# La valorisation économique des logiciels scientifiques

Nicolas Jullien (Nicolas.Jullien@imt-atlantique.fr), IMT Atlantique, LEGO-Marsouin

## 1    Introduction

Les établissements scientifiques, et leurs salariés, utilisent, adaptent, créent du logiciel. On pense aux outils métiers pour réaliser leur mission : gestion des enseignements (Moodle) ou support d'enseignement d'une matière (comme Maxima pour le calcul formel), par exemple. Nous parlerons ici des logiciels issus de travaux de recherche, conçus par un chercheur ou une équipe, dans le cadre d'un projet de recherche (financé, par l'ANR, l'Europe, etc. on non financé), ou en prestation de recherche pour un tiers.

Ces projets peuvent s'étendre sur des décennies (comme le projet d'assistant de preuve de programme Coq, ou la plateforme de distribution de contenu multimédia GPAC).

Nous discutons pourquoi ces logiciels sont produits, avec quelles ressources, l'intérêt que les établissements en retirent, ce que nous appellerons la « valorisation » des logiciels issus de la recherche scientifique. Celle-ci est multiple, comme le sont les missions des établissements scientifiques : valorisation sociale (contribution au patrimoine de connaissance mondiale), valorisation financière (contrats), valorisation économique (création d'entreprises), valorisation scientifique (publication), valorisation d'image (connaissance de l'institution parmi les publics-cibles : étudiants, chercheurs, entreprises, prescripteurs).

Les projets logiciels de l'activité de recherche scientifique peuvent s'articuler en trois catégories, qui se suivent temporellement, mais ne concernent pas tous les projets : preuve d'un concept scientifique, initiation d'une collaboration scientifique basée sur le logiciel, collaboration large sur la gestion d'un projet logiciel (dans la sphère de la recherche ou en tant qu'innovation c'est-à-dire qu'il répond à des besoins « industriels », ou « métier »).

Après avoir résumé ces grandes étapes, nous discuterons comment les institutions doivent s'organiser pour valoriser au mieux ces logiciels scientifiques.

## 2    Le logiciel comme preuve de concept

Au démarrage d'une production logicielle de recherche, il y a une idée scientifique qui a besoin d'être testée, « instanciée » dans une « preuve de concept scientifique », c'est-à-dire un programme qui permet de tester si l'idée scientifique fonctionne. Pour favoriser la réplicabilité des résultats, de plus en plus les sources de ce programme sont jointes à la publication, ce qui nécessite une licence (libre) l'autorisant.

Même si la production ne va pas plus loin, plusieurs valorisations sont possibles.

Si on connaît un partenaire d'être intéressé par cette production (éventuellement ce partenaire a passé commande du logiciel), et si le code appartient à l'institution, on peut organiser une négociation contractuelle de cession des droits, avec éventuellement un transfert de compétence sur le fonctionnement du logiciel. L'enjeu premier est financier et il s'agit de défendre au mieux les intérêts patrimoniaux et pécuniaires de l'institution.

S'il n'y a pas de partenaire identifié, il peut être intéressant de publier ou de continuer à publier sous une licence libre les codes de bonne qualité, afin d'illustrer les compétences de l'institution et faciliter la diffusion des publications scientifiques réalisées avec. C'est aussi la seule manière d'identifier un partenaire intéressé par une collaboration de transfert. Il faut cependant veiller à bien préciser qu'il s'agit d'un logiciel non maintenu.

Si ces premiers résultats ouvrent de nouvelles perspectives, ils peuvent être étendus (les concepts sont approfondis, développés dans des preuves de concept supplémentaires), consolidés (le code est amélioré, le logiciel est rendu plus solide pour être réellement utilisé). Si les contributions restent internes à l'équipe, à l'institution, la gestion de ces contributions peut rester informelle, notamment sur les questions de propriété du code (même s'il est bon de mettre en place au plus tôt une traçabilité des contributions). D'une simple preuve de concept, le développement est devenu un projet qui fédère l'équipe, qui peut devenir central pour le laboratoire.

La publication sous licence libre du logiciel permet que ces extensions d'usage ou de recherche soient publiées par d'autres que les auteurs initiaux, ce qui, peut-être de façon contre intuitive, favorise la valorisation.

# 3     Le logiciel de collaboration scientifique

Certains logiciels diffusés vont être repris par d'autres équipes, qui peuvent faire des contributions mineures (bug report), ou proposer des évolutions (spécification ou commit en code) permettant petit à petit d'étendre les fonctionnalité du logiciel. Cette amélioration collaborative incrémentale par des utilisateurs développeurs est à la base du concept de logiciel libre[1]. C'est aussi le fonctionnement de la recherche et ce qui justifie l'idée de science ouverte[2].

Le producteur de la recherche (ici le logiciel) est ainsi incité à publier cette recherche pour obtenir des retours, voire même à des innovations complémentaires et cumulatives. En miroir, pour que les retours soient validés et intégrés dans la version officielle du logiciel, les contributeurs externes ont intérêt à les soumettre au projet initial, qui accélère le développement de sa propre recherche ou de son innovation, devenant encore plus attractivité. Si c'est le cas, l'équipe initiale doit s'organiser pour gérer ces demandes et faire des retours à ces/ses utilisateurs avancés.

Un code sous licence libre illustrant une idée scientifique est devenu ce que nous appellerons un « logiciels de collaboration scientifique ». La valorisation reste scientifique (publication, citation des utilisateurs, etc.) La visibilité du logiciel dans la communauté scientifique, qui assure la renommée de l'équipe de recherche et de l'institution, signale son expertise. Si ce logiciel peut servir de démonstrateur à des travaux scientifiques, il facilite l'accès à des contrats (projets européens où ces plateformes sont très importantes, contrats de recherche industriels). La visibilité du projet renforce celle du laboratoire, et renforce son attractivité pour les chercheurs qui peuvent vouloir le rejoindre (doctorants et permanents).

Plus le projet est important sur le plan du volume de code, ou de participants, plus les contraintes techniques et organisationnelles deviennent fortes. Il s'agit d'organiser, de gérer ce qui a été appelé une « architecture de collaboration »[3].

# 4     Le projet logiciel

Développer une plateforme qui permet de gérer des contributions multiples, externes et internes à l'organisation initiale demande des investissements, qui dépassent le cadre des compétences scientifiques.

---

[1] Voir le texte de Richard Stallman, sur les fondements du logiciel libre, et Hippel, Eric von, and Georg von Krogh. "Open source software and the "private-collective" innovation model: Issues for organization science." *Organization science* 14.2 (2003): 209-223.

[2] Cohen, Wesley M., and Daniel A. Levinthal. "Absorptive capacity: A new perspective on learning and innovation." *Administrative science quarterly* (1990): 128-152.

[3] West, Joel, and Siobhán O'mahony. "The role of participation architecture in growing sponsored open source communities." *Industry and innovation* 15.2 (2008): 145-168.

Il faut souvent ré-écrie le code, autour d'un cœur, qui assure la base du service et de l'innovation et des modules périphériques, qui instancient la variété des usages possibles. Cela afin de faciliter l'entrée de nouveaux contributeurs, sur des mises en applications spécifiques de l'innovation, tout en permettant aux porteurs du projet de se concentrer sur le cœur du logiciel, racine de l'innovation.

Pour ce faire, il faut aussi structurer les règles de contribution, les outils techniques d'évaluation et d'acceptation du code soumis. Ce travail d'édition du logiciel : gestion technologique, gestion des demandes, publication des versions, demande des compétences spécifique, s'éloigne du métier de chercheur. Il s'agit d'organiser la gestion de ce qu'on peut appeler un « projet de développement logiciel (scientifique) », ou plus simple « projet logiciel » sur plusieurs années.

Si le projet reste stratégique pour l'institution, c'est-à-dire que la composante recherche reste forte, il va falloir qu'elle finance ces ressources ; il s'agit d'affecter une ou plusieurs personnes sur la gestion du projet, personnes qui doivent avoir la double compétence, rare, de gestionnaire de projet informatique et chercheur, comprenant les fondements scientifiques de projet. Outre l'importance pour la recherche, ces succès renforcent l'attractivité en tant que plateforme évoquée plus haut, mais aussi l'attractivité des producteurs (équipe et institution), aussi bien auprès des centres de recherche, que des entreprises ou des étudiants (un logiciel est une réalisation que l'on peut « montrer »).

Si les développements liés à la recherche diminuent, et que les usages et les contributions sont de plus en plus liés à des questions d'usages « industriels », l'institution de recherche n'est pas forcément la mieux placée pour gérer le projet : ce n'est pas le métier des chercheurs, et plus généralement les institutions de recherche ne sont sans doute pas équipées pour gérer l'innovation (la recherche et développement de nouveaux usages), ou la fourniture de prestations contractuelles de support à l'utilisation du logiciel.

Il s'agit alors pour l'institution d'assurer ce transfert, une transition des usages de recherche vers des usages plus industriels, de récupérer éventuellement une valorisation financière, mais aussi de participer à la mission de développement économique par la création d'activités industrielles.

La séparation de l'équipe initiale (et de l'institution propriétaire du logiciel) se fait souvent à ce moment-là, en confiant le travail d'édition du logiciel à un tiers : soit une fondation, soit un éditeur privé (start-up issue de l'équipe initiale, par exemple), qui monétise les demandes d'évolution ou/et d'expertise pour financer les développeurs nécessaire pour gérer le projet techniquement et l'animation de la communauté. L'institution pourra être un des clients de la start-up (maintenance, développement spécifique) pour assurer son démarrage. Il peut y avoir une prise de participation dans la start-up pour garder une certaine autorité sur les actions de la start-up vis-à-vis des articulations avec la recherche.

Les possibilités de transfert dépendent aussi de qui est propriétaire du code, de comment le développement du projet logiciel et l'intégration des différentes contributions a été gérée dans les phases précédentes.

Cela nous amène à proposer quelques recommandations auprès des institutions de recherche pour la valorisation des logiciels issus de la recherche.

# 5 Des valorisations à articuler

On le voit, les termes de « protection », « valorisation » et « publication », « science ouverte » et « logiciel libre » ne sont pas forcément antagonistes. Mais les articuler demande d'être capable d'évaluer la portée du logiciel à valoriser, au regard des efforts qu'on est capable d'y mettre. C'est une analyse à mener collaborativement, entre les chercheurs, les mieux placés pour évaluer le po-tentiel (scientifique) du projet, et l'institution, qui doit évaluer les ressources qu'elle peut y affecter.

Car maintenir un logiciel est coûteux en temps-homme, surtout si ce logiciel attire de nombreuses contributions. Le métier des institutions publiques n'est pas non plus d'assurer une diffusion commerciale des logiciels, ou des services d'aide à l'utilisation.

Il est difficile d'anticiper le succès d'un code, de savoir quel algorithme développé et publié pour accompagner une publication va devenir un projet de logiciel, gérant une plateforme mondialement connue. Ces évolutions prennent du temps, et passent par différentes étapes où les chercheurs, le laboratoire, l'institution qui les ont initiées peuvent réviser l'intérêt du projet et de leur investissement.

Deux recommandations nous semblent importantes pour s'assurer de pouvoir gérer les différentes valorisations possibles.

D'abord on parle ici de logiciels scientifiques, forcément créés par des laboratoires. C'est dans ces laboratoires qu'ils ont le plus de chance de croître en développant d'abord leur impact scientifique, et ensuite, éventuellement leur impact économique. Mais le métier de gestion d'un projet logiciel est différent de celui actuellement reconnu dans la carrière d'un chercheur. Permettre à certains chercheurs (post-doc ou permanent) de consacrer du temps à cette maintenance, ou favoriser, dans les laboratoires gros producteurs de logiciels, la localisation d'ingénieurs de recherche qui ont cette double compétence scientifique et en génie logiciel et un préalable à la valorisation de plus de logiciels scientifiques.

Il faut aussi s'assurer très tôt dans le projet de maîtriser la propriété liée au code développé. Cela veut dire :
- savoir ce qu'ont développé les différents contributeurs, et avec quel statut (salariés de l'institution, apports temporaires de non salariés --stages, thèses de doctorat--, contributions externes) ;
- généraliser et automatiser les contrats de cessions de droit pour les contributeurs non salariés, en échange d'une publication sous une licence libre par l'unique propriétaire, l'institution porteuse, par exemple.

Cela nous semble la meilleure façon de favoriser et une diffusion dans le cadre de la science ouverte, et une maîtrise, sur le long terme, des différentes valorisations possibles. C'est facile à mettre en place sur une plateforme en ligne de développement logiciel (« forge »), privée (comme GitHub ou GitLab) ou publique (au niveau des institutions ou mutualisée à l'image de Hal pour les publications scientifiques).

Développer ces habitudes passera par la formation des chercheurs et des services de valorisation à la production du logiciel et aux différentes phases de sa valorisation.